# A New Reversible TSG Gate and Its Application For Designing Efficient Adder Circuits

Himanshu Thapliyal
Center for VLSI and Embedded System Technologies
International Institute of Information Technology
Hyderabad-500019, India
(thapliyalhimanshu@yahoo.com)

M.B Srinivas
Center for VLSI and Embedded System Technologies
International Institute of Information Technology
Hyderabad-500019, India
(srinivas@iiit.net)

*Abstract*— **In the recent years, reversible logic has emerged as a promising technology having its applications in low power CMOS, quantum computing, nanotechnology, and optical computing. The classical set of gates such as AND, OR, and EXOR are not reversible. This paper proposes a new 4 * 4 reversible gate called "TSG" gate. The proposed gate is used to design efficient adder units. The most significant aspect of the proposed gate is that it can work singly as a reversible full adder i.e reversible full adder can now be implemented with a single gate only. The proposed gate is then used to design reversible ripple carry and carry skip adders. It is demonstrated that the adder architectures designed using the proposed gate are much better and optimized, compared to their existing counterparts in literature; in terms of number of reversible gates and garbage outputs. Thus, this paper provides the initial threshold to building of more complex system which can execute more complicated operations using reversible logic.**

## I. INTRODUCTION

Researchers like Landauer have shown that for irreversible logic computations, each bit of information lost, generates kTlog2 joules of heat energy, where k is Boltzmann's constant and T the absolute temperature at which computation is performed [1]. Bennett showed that kTln2 energy dissipation would not occur, if a computation is carried out in a reversible way [2], since the amount of energy dissipated in a system bears a direct relationship to the number of bits erased during computation. Reversible circuits are those circuits that do not lose information. Reversible computation in a system can be performed only when the system comprises of reversible gates. These circuits can generate unique output vector from each input vector, and vice versa. In the reversible circuits, there is a one-to-one mapping between input and output vectors. Bennett's theorem [2] about heat dissipation is only a necessary and not sufficient condition, but its extreme importance lies in the fact that every future technology will have to use reversible gates to reduce power. As the Moore's law continues to hold, the processing power doubles every 18 months. The current irreversible technologies will dissipate a lot of heat and can reduce the life of the circuit. The reversible logic operations do not erase (lose) information and dissipate very less heat. Thus, reversible logic is likely to be in demand in high speed power aware circuits. Reversible circuits are of high interest in low-power CMOS design, optical computing, nanotechnology and quantum computing. The most prominent application of reversible logic lies in quantum computers. A quantum computer will be viewed as a quantum network (or a family of quantum networks) composed of quantum logic gates; each gate performing an elementary unitary operation on one, two or more two–state quantum systems called qubits. Each qubit represents an elementary unit of information; corresponding to the classical bit values 0 and 1. Any unitary operation is reversible hence quantum networks effecting elementary arithmetic operations such as addition, multiplication and exponentiation cannot be directly deduced from their classical Boolean counterparts (classical logic gates such as AND or OR are clearly irreversible).Thus, Quantum Arithmetic must be built from reversible logical components [10].

One of the main constraints in reversible logic is to minimize the number of reversible gates used and garbage output produced. Garbage output refers to the output that is not used for further computations. In other words, it is not used as a primary output or as an input to other gate. In literature, there are a number of existing reversible gates such as Fredkin gate[3,4,5],Toffoli Gate[3,4] and New Gate [6]. In this paper, the focus is on the proposal of new reversible 4*4 TSG gate. The proposed TSG gate is used to design optimized adder architectures like ripple carry adder and carry skip adder. It has been proved that the adder architectures using the proposed TSG gate are better than the existing one in literature; in terms of number of reversible gates and garbage outputs. Thus, the paper provides the initial threshold to build more complex systems which can which can execute more complicated operations. The reversible circuits designed and proposed here form the basis of the ALU of a primitive quantum CPU[11,12,13].

## II. PROPOSED 4*4 REVERSIBLE GATE

In this paper, a 4 * 4 one through reversible gate called TS gate "TSG" is proposed. The proposed reversible TSG gate is shown in Fig. 1. The corresponding truth table of the gate is shown in Table I. It can be verified from the Truth Table that the input pattern corresponding to a particular output pattern can be uniquely determined. The proposed TSG gate can implement all Boolean functions. Fig. 3 shows the implementation of the proposed gate as XOR function. Fig. 2 and Fig. 4 shows the implementation of the proposed gate as NOT and NOR function respectively.

TABLE I. TRUTH TABLE OF THE PROPOSED TSG GATE

| A | B | C | D | P | Q | R | S |
|---|---|---|---|---|---|---|---|
| 0 | 0 | 0 | 0 | 0 | 0 | 0 | 0 |
| 0 | 0 | 0 | 1 | 0 | 0 | 1 | 0 |
| 0 | 0 | 1 | 0 | 0 | 1 | 1 | 1 |
| 0 | 0 | 1 | 1 | 0 | 1 | 0 | 0 |
| 0 | 1 | 0 | 0 | 0 | 1 | 1 | 0 |
| 0 | 1 | 0 | 1 | 0 | 1 | 0 | 1 |
| 0 | 1 | 1 | 0 | 0 | 0 | 0 | 1 |
| 0 | 1 | 1 | 1 | 0 | 0 | 1 | 1 |
| 1 | 0 | 0 | 0 | 1 | 1 | 1 | 0 |
| 1 | 0 | 0 | 1 | 1 | 1 | 0 | 1 |
| 1 | 0 | 1 | 0 | 1 | 1 | 1 | 1 |
| 1 | 0 | 1 | 1 | 1 | 1 | 0 | 0 |
| 1 | 1 | 0 | 0 | 1 | 0 | 0 | 1 |
| 1 | 1 | 0 | 1 | 1 | 0 | 1 | 1 |
| 1 | 1 | 1 | 0 | 1 | 0 | 0 | 0 |
| 1 | 1 | 1 | 1 | 1 | 0 | 1 | 0 |

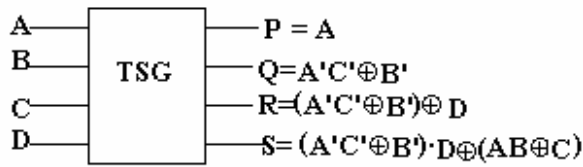

Figure 1. Proposed Reversible TSG Gate

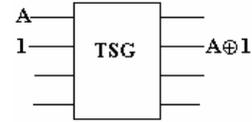

Figure 2. Proposed TSG Gate As NOT Gate

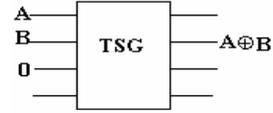

Figure 3. Proposed TSG gate as XOR Gate

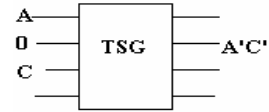

Figure 4. Proposed TSG Gate as NOR Gate

Since, the NOR gate is a universal gate and any Boolean function can be implemented through it. Hence, the proposed gate can be used to implement any Boolean function. One of the prominent functionality of the proposed gate is that that it can work singly as a reversible Full adder unit. Fig. 5 shows the implementation of the proposed gate as the reversible Full adder.

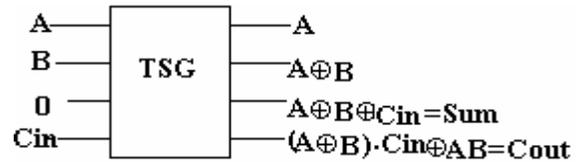

Figure 5. Proposed TSG Gate Working Singly As Reversible Full Adder

A number of reversible full adders were proposed in [6,7,8,9,14]. The reversible full adder circuit in [6] requires three reversible gates (two 3*3 new gate and one 2*2 Feynman gate) and produces three garbage outputs. The reversible full adder circuit in [7,8] requires three reversible gates (one 3*3 new gate, one 3*3 Toffoli gate and one 2*2 Feynman gate) and produces two garbage outputs. The design in [9] requires five reversible Fredkin gate and produces five garbage outputs. The design in [14] requires two reversible gates (one 3*3 New gate and one 3*3 New Toffoli Gate) and produces two garbage outputs. The proposed full adder using TSG in Fig. 5 requires only one reversible gate (one TSG gate) and produces only two garbage outputs. Hence, the full-adder design in Fig. 5 using TSG gate is better than the previous full-adder designs of [6,7,8,9,14]. A comparative result is shown in Table II.

TABLE II. COMPARITIVE RESULTS OF DIFFERENT FULL ADDER CIRCUITS

|  | No of Gates | No of Garbage Outputs |
|---|---|---|
| Proposed Circuit | 1 | 2 |
| Existing Circuit[6] | 3 | 3 |
| Existing Circuit [7,8] | 3 | 2 |
| Existing Circuit[9] | 5 | 5 |
| Existing Circuit[14] | 2 | 2 |

## III. APPLICATIONS OF THE PROPOSED TSG GATE

To illustrate the application of the proposed TSG gate, two different types of adders – ripple carry and carry skip adders are designed. It has been proved that the adders circuit drawn using the proposed gate are the most optimized one compared to their existing counterparts in literature.

### A. Ripple Carry Adder

The full adder is the basic building block in the ripple carry adder. The full adder circuit using the proposed TSG gate is shown in Fig. 6. The ripple carry adder is obtained by cascading the full adders in series. The output expressions for a ripple carry adder are :
$S_i = A \oplus B \oplus C_i$;
$C_{i+1} = (A \oplus B) \cdot C_i \oplus AB$  (i=0,1,2….)

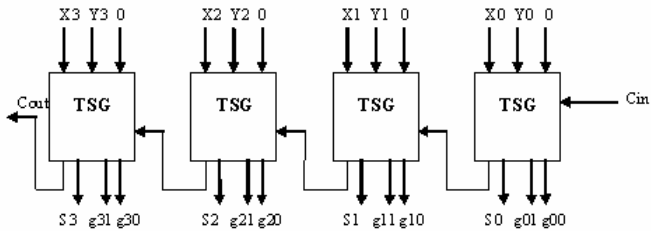

Figure 6. Ripple Carry Adder Using The Proposed TSG Gate

#### 1) Evaluation of the Proposed Ripple Carry Adder

It can be inferred, from the Fig. 6 that for N bit addition; the proposed ripple carry adder architecture uses only N reversible gates and produces only 2N garbage outputs. There also exists previously proposed ripple carry adders in the literature. But, the ripple carry adder using our proposed TSG gate is the most optimized one. Table III shows the result that compares the proposed ripple carry adder using TSG gate, with the existing full adders of [6,7,8,9,14]. It is observed that the proposed circuit is better than the existing circuits; both in number of reversible gates and garbage outputs.

TABLE III. COMPARITIVE RESULTS OF DIFFERENT RIPPLE CARRY ADDER CIRCUITS

|  | No of Gates | No of Garbage Outputs |
|---|---|---|
| Proposed Circuit | N | 2N |
| Existing Circuit [6] | 3N | 3N |
| Existing Circuit[7,8] | 3N | 2N |
| Existing Circuit[9] | 5N | 5N |
| Existing Circuit[14] | 2N | 2N |

### B. Carry Skip Adder

In the carry skip adder, delay is reduced due to the carry computation. In the full adder operation, if either input is a logical one, the cell will propagate the carry input to its carry output. Hence, the ith full adder carry input $C_i$, will propagate to its carry output, $C_{i+1}$, when $P_i = X_i \oplus Y_i$. In addition, the multiple full adders, making a block can generate a "block" propagate signal P to detour the incoming carry around to the block's carry output signal. Fig. 7 shows a four bit carry skip adder block. It is quickly determined by each block, that whether the block's carry input is propagated to its carry output. If the block propagate P is one, the block carry input Cin is propagated as the block carry output Cout. An AND gate is used to generate a block propagate signal P. Fig. 8 shows the proposed carry skip compatible Full adder constructed with TSG gate.

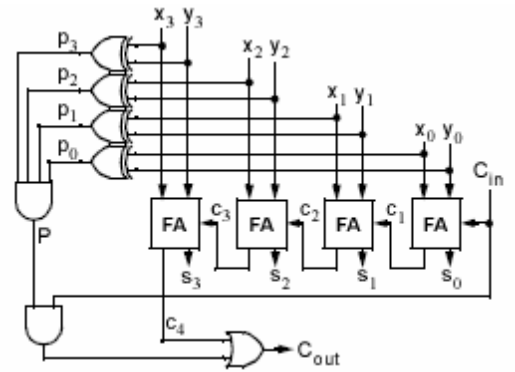

Figure 7. Four Bit Carry Skip Adder Block

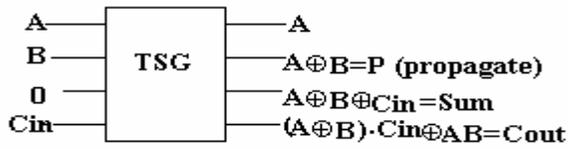

Figure 8. Full adder Circuit (With Propagate) Using TSG Gate

The conventional skip block in Fig. 7 uses the AND-OR gate combination. Fig. 9 shows the block diagram of the carry skip adder block constructed with TSG gates and Fredkin gates (FG). The three FGs in the middle of Fig. 9 are used to perform the AND4 operation. This will generate the block propagate signal P. The single FG in the left side of Fig. 9 performs the AND-OR function to create the carry skip logic and generate the block carry out signal Cout. In the proposed carry Skip adder, the FG propagates the block's carry input to the next block if the block propagate signal P is one; otherwise, the most significant full adder carry C4 is propagated to the next block. The traditional carry skip AND-OR logic in Fig. 7 and the carry skip logic in Fig. 9 do not have the equivalent truth Tables but it must be noted that the Fredkin carry skip logic more faithfully adheres to the spirit of carry skip addition by propagating the correct value of Cin to Cout. The FG carry skip logic in Fig. 9 can lead to improve carry propagation, when the block carry propagate signal P is one.

When P is one, the block carry input Cin must propagate to the next block, independent of the result of the carry C4 created by the full adders within the block. When P=0, the traditional AND-OR carry skip logic in Fig. 9 must account for sending C4, it does not perform its carry skip operation efficiently. The Fredkin carry skip logic in Fig. 9 passes Cin to Cout whenever P=1, regardless of C4. The savings in time can be quite significant as block sizes increases. The proposed N bit carry skip adder requires N TSG gates for implementation of ripple carry adder. Further, the N input AND gate will require N-1 FGs. Thus the total gates required in proposed N bit carry save adder is 2N. Table IV shows the comparative results of reversible carry skip adders.

TABLE IV.   COMPARATIVE RESULTS OF DIFFERENT REVERSIBLE CARRY SKIP ADDER CIRCUITS

|  | No of Gates | No of Garbage Outputs |
|---|---|---|
| Proposed Circuit | 2N | 3N |
| Existing Circuit[9] | 6N | 12N |

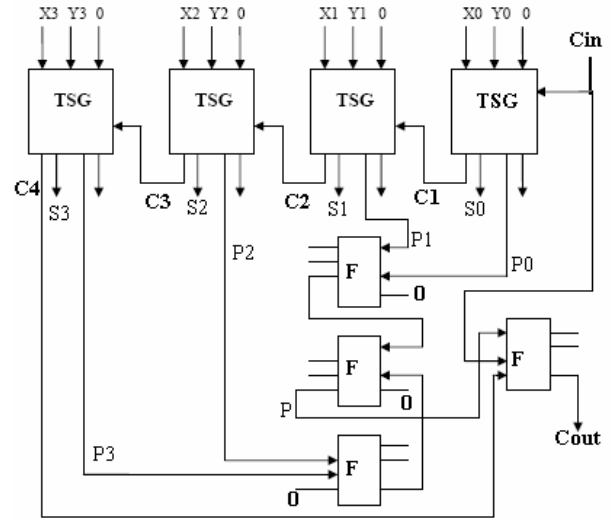

Figure 9. Four Bit Carry Skip Adder Block Using Proposed TSG and Fredkin (F) gates

IV. CONCLUSIONS

The focus of this paper is the proposal of new reversible 4*4 TSG gate. The proposed TSG gate is being used to design optimized architectures of ripple carry and carry skip adders. It is proved that the adder architectures using the proposed TSG gate are better than the existing counterparts in literature, in terms of number of reversible gates and garbage outputs. All the proposed architectures are analyzed in terms of technology independent implementations. The technology independent analysis is necessary since quantum or optical logic implementations are not available. There are a number of significant applications of reversible logics such as low power CMOS, quantum computing, nano-technology, and optical computing and the proposed TSG gate and efficient adder architectures are one of the contributions to reversible logic. The proposed circuit can be used to design large reversible systems. In a nutshell, the advent of reversible logic will significantly contribute in reducing the power consumption. Thus, the paper provides the initial threshold to build more complex systems which can which can execute more complicated operations. The reversible circuits designed and proposed here form the basis of the ALU of a primitive quantum CPU.